\documentclass[letter]{jpsj2}

\def\runauthor{Toshifumi Itakura}

\title{
Effect of multiple charge traps on dephasing rates \\
of a Josephson charge qubit system
}

\author{%
Toshifumi Itakura and Yasuhiro Tokura
\thanks{E-mail address: itakura@will.brl.ntt.co.jp}
}

\inst{%
   NTT Basic Research Laboratories, NTT corporation \\
  3-1, Morinosato Wakamiya, Atsugi-shi,
  Kanagawa Pref., 243-0198 Japan \\
    Tel : +81-46-240-3938
    Fax : +81-46-240-4726 }

\recdate{\today}

\abst{%
We examine the dephasing rate of a Josephson charge qubit system
    due to background charge fluctuations.
We consider single qubit and two-charge traps.
The transition probability was controlled to 
  a state where two traps were occupied.
The transition probability was  affected by the
  Coulomb blockade effect that occurs between two charge traps.
To obtain the dephasing rate, we 
   computed the spectra of random
   frequency modulation signals.  
Our results show that the interaction between charge traps
   suppresses dephasing.}

\kword{%
Josephson charge qubit, quantum computation, dephasing, 
background charge fluctuations 
}

\begin{document}
\maketitle

Among various proposals, quantum bits (qubits) in solid state materials,
           such as, superconducting Josephson junctions
\cite{Nakamura}
           and quantum dots
\cite{Tanamoto,Miranowicz,Loss},
           have the advantage of  scalability.
Proposals to implement a quantum computer using superconducting nanocircuits
    are proving to be very promising
    \cite{Makhlin,Averin}
    and several experiments have already highlighted the quantum properties
    of these devices
    \cite{Bouchiat}.
Such a coherent-two-level system constitutes  a qubit and
 the quantum computation can be carried out as the
        unitary operation functioning on the multiple qubit system.
Essentially, this quantum coherence must be maintained during computation.
However, dephasing is hard to 
     avoid due to the system's interaction with
     the environment.
In terms of a bonding-antibonding bases,
            the decay of off-diagonal elements of
            the qubit density matrix signals that dephasing is occurring.
This dephasing is characterized by the dephasing time $T_2$.
Various environments can cause dephasing.
In superconducting nanocircuits various sources of decoherence
     are present
     \cite{Makhlin},
     such as fluctuations originating from the surrounding circuit,
     quasiparticle tunneling,
     background charge fluctuation (BCF),
     and flux noise.
For a charge qubit system, BCF is 
            one of the most critical dephasing channels
\cite{Itakura_Tokura,Nakamura_CE,Fazio,Shnirman}.

BCFs 
         have been observed in various kinds of  systems
\cite{Devoret,Martinis,Lyon,Zorin}.
In nanoscale systems, they are
     the electrostatic potential fluctuations
     due to the dynamics of electrons, or holes
     on a charge trap.
In particular, the charge at a charge trap 
     fluctuates
     with the  Lorentzian spectrum form,
     which is 
     called
     random telegraph noise
     in the time domain
\cite{Lyon,Fujisawa_BC}.     
The random 
      distribution of the positions of such dynamical charge traps
       and their time constants  
     lead to  BCFs or 1/f noise
\cite{BC}.
In solid-state charge qubits,
    these BCFs
    result in  a  dynamical electrostatic disturbance and
    hence, dephasing.
The theoretical effect of 1/f noise on a charge Josephson
     qubit has been examined previously 
     \cite{Itakura_Tokura,Nakamura_CE,Fazio,Shnirman}.
    
We 
       investigated 
       how the electrostatic disturbance
      coming from
      two or more dynamical charge traps
      affects the quantum coherence of a qubit.
In past studies, an environment composed of free charge traps had been 
      considered
      \cite{Itakura_Tokura,Fazio}.
When such an environment is interacting with itself,
      its characteristic nature would be expected to 
      affect the relaxation phenomena.       
In present study, 
   we especially concentrated on the correlation effect between the charges
     in the traps.
We consider  pure dephasing as an event which
     occurs when the dynamical charge traps
     induce fluctuation in extra bistable bias. 
It should be noted that this dephasing process does not mean 
        the qubit 
        is entangled with the environment,
        but rather,
        that  
        the stochastical evolution of an external classical field,
        is suppressing the off-diagonal density matrix elements 
        of the qubit after being averaged out over statistically distributed
         samples.

The system under consideration is Cooper pair box
     \cite{Makhlin}.
Under appropriate conditions ( charging energy $E_C$ much larger than the
     Josephson coupling $E_J$ and temperatures $k_B T \ll E_J$)
     only two charge states are important,
     and the Hamiltonian of the qubit $H_{qb}$ reads      
\begin{equation}
  H_{qb} = \frac{\delta E_C}{2} \sigma_z  +
             \frac{E_J}{2} \sigma_x  
\end{equation}
where  the charge bases $\{|0>,|1>\}$ is expressed using the Pauli matrices,
       and the bias $\delta E_C \equiv E_C ( 1 - C_x V_x/e)$
       can be turned by varying the applied gate voltage $V_x$.
The environment is a set of BCF electrostatistically coupled to the qubit
\cite{Itakura_Tokura,Fazio,Galperin,Nazarov},
\begin{equation}             
  H_{qb-imp} = \sum_{i=1}^N 
  \frac{ \hbar J_{Ci} }{2} \sigma_z 
  ( d^{\dagger}_i d_i - \frac{1}{2} )
\end{equation}
        where $d^{\dagger}_i$ and $d_i$ are 
        the electron creation and annihilation operators
        of a charge trap,
        $i$ is the index of  N charge traps,
        and the coupling with the qubit is such that each BCF produces
        a bistable extra bias 
         $\hbar J_{Ci}$.
Because qubit Hamiltonian consists of $E_J$ and $\delta E_C$,
        the dephasing consists of that with dissipation
        and pure dephasing.
In general,  the dephasing with dissipation can be neglected
        as follows.         
For physical setups,
   $\delta E_C \simeq 122$ $\mu$eV, and $E_J \simeq 34$ $\mu$eV
\cite{Nakamura_CE};
By perturbation method,
        \cite{Makhlin}
        the ratio of the dephasing rate with  dissipation to
        pure dephasing rate 
        is roughly  given by
        $\frac{E_J^2}{\delta E_C^2} 
        \frac{\lambda^2}{(\delta E_C^2 +E_J^2)/\hbar^2 + \lambda^2}$
        in the presence of 
        the bistable extra bias,
        where $\lambda$ is the transition rate of 
        the dynamical charge trap. 
For the above experimental setups
        with the dynamical charge trap with low frequency,
        we can neglect the effect of $E_J$ because $E_J < \delta E_C$ and
        $\frac{\sqrt{\delta E_C^2 + E_J^2}}{\hbar} \gg \lambda$.
Then the pure dephasing event is critical.
In final results of present study, we discuss about the
        many charge traps which are interacting with each other.
For this case, the dominant process is different,
        we discuss about this behavior latter. 
We neglect the back action from the qubit to charge traps,
        namely, the transition rates of charge traps
        do not depend on the qubit state.
This assumption is justified because the qubit induces {\it static}
        shift of
        the chemical potential of  charge traps.
And the change in the chemical potential
        of a charge trap does not renormalize the
        transition rate of itself.
\cite{Haldane}
        
Using the environment variable  
        $X_i (t)  (=<d_i^{\dagger} (t) d_i (t) >_r - 1/2 )$,
        where $<A(t)>_r$ is trace 
         of the operator $A(t)$ 
        about the electron reservoir of the charge trap, 
        we rewrite
        the Hamiltonian 
        in terms of the  Pauli matrix as
\begin{equation}
{\cal H} =
\frac{ \delta E_C}{2} \sigma_z + 
          \sum_i^N \frac{\hbar J_{Ci}}{2} \sigma_z X_i (t),
\end{equation}
we assume that the charge traps are strongly coupled with
            their charge reservoirs
            and the time evolution of $ X_i(t)$
            is a Poisson process.              

Following the time evolution of density matrix of qubit,
   we obtain the following off-diagonal element 
$   \rho_{12} (t) = \rho_{12} (t_0) e^{i \delta E_C / \hbar ( t - t_0 )}
   e^{ i \int_{t_0}^t d \tau x (\tau)},$
    where $t_0$ is the initial time, and
    $ x (t) = \sum_i^N J_{Ci} X_i (t) $
    takes $2^N$ with possible different values of $a_1, \dots, a_{2^N}$.
The fluctuation in tunneling coupling constant is pure dephasing and
    does not accompany  relaxation of the population.
Therefore, diagonal elements of the qubit density matrix do not change.

In the following, we estimate the ensemble average 
    of off diagonal element of the density matrix, 
 $   E[ \rho_{12} (t)] = \rho_{12} (t_0) e^{ i \delta E_C / \hbar ( t -t_0 )}
            < e^{ i \int_{t_0}^t d \tau x (\tau) }>.$
For this quantity, we can apply the characteristic functional method
\cite{Kubo,Papoulis},
 namely,
 $    R ( t )  \equiv  < e^{ - i \int_0^t d \tau x ( \tau ) } > 
              =  \sum_{i,k=1}^{2^N} p_i R_{ik} (t),$
where $p_i$ is the occupation probability of the state $i$,
which can be determined by the stationary condition,
 $ 0 = - \mu_i p_i (t) + \sum_{j=1, \ne i}^{2^N} \lambda_{ji} p_j (t),$
where $\lambda_{ij}$ is the transition probability  defined
for very short time $\Delta t$ with $i \ne j$,
and $\mu_{i} \equiv \sum_{j \ne i} \lambda_{ij}$
 is the emission rate during this time.
The $p_i$ has the properties
$\sum_i p_i = 1$
and
$ \mu_i p_i = \sum_{j \ne i} \lambda_{ji} p_j$.
The average of $x$, $\eta$, is given by $\sum_i p_i a_i$
and the variance $\sigma$ is given by $\sqrt{\sum_i p_i a_i^2 - \eta^2}$.
    The function $R_{ik}$ satisfies following
    the first order differential equation,
\begin{eqnarray}
          \label{eqn:de}
     \frac{d R_{ik} (t) }{dt} &=& ( i a_k - \mu_k ) R_{ik} (t)
     + \sum_{m \ne k} \lambda_{mk} R_{km} (t) \nonumber \\
      & \equiv &\sum_{m} R_{im} \Lambda_{mk},
\end{eqnarray}
with the initial condition $ R_{ij} (t=0) = \delta_{ij}$.
$T_2^{-1}$ characterizes the exponential tail of long-time dephasing behavior.
This quantity is obtained by Min$(- {\rm Re} (\epsilon_i) )$,
    where $\epsilon_i$'s are the eigenvalues of $\Lambda$.
While for very short $t$,
 the $R(t)$ shows Gaussian behavior 
 \cite{Itakura_Tokura}.

First, we examine the single charge trap case (N=1)
where we have
\begin{equation}
         \Lambda =\left(
         \begin{array}{cc}
         i a_1 - \lambda_u  & \lambda_u \\
         \lambda_d & i a_2 - \lambda_d
         \end{array}
         \right) .
\end{equation}
where the $\lambda_u$ ($\lambda_d$) is the transition rate from
 the 1st state to the 2nd state ( 2nd state to the 1st state).
The resultant $T_2^{-1}$ is given by
$     T_2^{-1} = \frac{1}{2} ( \lambda_u + \lambda_d - {\rm Re} \sqrt{A})
$
where $A=4a_1 a_2 + 4 i ( a_1 \lambda_d + a_2 \lambda_u )
  + ( \lambda_u + \lambda_d - i a_1 - i a_2 )^2$, and $J_C=a_2 - a_1$.
For weak coupling ($|J_C| \ll  {\rm max} ( \lambda_{u}, \lambda_{d}$)),
    $J_C=d=-2a_1=2a_2$,
    and $d$ characterize strength of bistable extra bias,
    $T_2^{-1}$ is given by
\begin{equation}
    \label{eqn:T2weak}
    T_2^{-1} = \lambda_u \lambda_d d^2 /( \lambda_u + \lambda_d )^3.
\end{equation}
For strong coupling, ($|d| \gg \lambda_{u,d}$),
    $T_2^{-1} = ( \lambda_u + \lambda_d )/2$.           
These results coincide with those found by Itakura and Tokura
 \cite{Itakura_Tokura},
      where the dephasing time was 
      derived using a different method.

Next, we examined the two traps, (N=2)
    including the Coulomb blockade effect occurring between the traps.
When two traps are located
    close to one other,
    there should be  capacitance coupling between
 two occupied traps.
However, we neglect tunneling between the charge traps.
There are four states: both charge traps empty,   left charge trap occupied, 
    right charge trap occupied, and both charge traps occupied (Fig. 1).
$\lambda_{ij}'s$ are transition rates from $i$ state to $j$ state.
and we neglect the transition processes between 1 state and 4 state, 
  and 2 state and 3 state.
In general, we notice $\lambda_{12} \geq \lambda_{34}$ 
and $\lambda_{13} \geq \lambda_{24}$,
where the equations are for the absence of Coulomb blockade effect.

\begin{figure}
\center
\includegraphics[width=.5\linewidth]{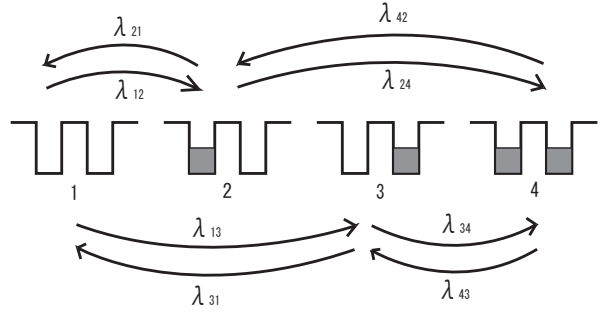}
\begin{minipage}[t]{8.5cm}
\caption{Scheme of two charge trap system.
The wells represent the charge traps and
the arrows represent the transition between each of four states.
$\lambda_{ij}$'s indicate the transition probabilities
from state $i$ to state $j$. 
}
\label{figure1}
\end{minipage}
\end{figure}

For actual calculation,
 we restricted the parameters for the transition rate  
    which are symmetric for two charge traps,
    $\lambda_{u} = \lambda_{12} = \lambda_{13},
     \lambda_{d} = \lambda_{21} = \lambda_{31},
    \lambda_{u}' = \lambda_{24} = \lambda_{34},
    \lambda_{d}' = \lambda_{42} = \lambda_{43}$.
The occupation probabilities are
    $ p_1 = \frac{\lambda_{d} \lambda_{d}'}{D}, 
    p_2 = p_3 =\frac{\lambda_{u} \lambda_{d}'}{D},
    p_4 = \frac{\lambda_{u} \lambda_{u}'}{D}$, where
    $ D = \lambda_{d} \lambda_{d}' + ( 2 \lambda_{d}'+ \lambda_{u}'
    ) \lambda_{u}$. 

First we discuss the high temperature behavior,
  where energy differences of each state
  are lower than temperatures,
  while $\delta E_C$ and gap energy of
  cooper pair of qubit are much higher than temperatures.         
To demonstrate the effect of Coulomb interaction transparently,
   we chose parameters, 
   $\lambda_u = \lambda_d = \lambda_d' \equiv \lambda$ and 
   $\lambda_u' \equiv \lambda'$,
   and calculate $T_2$ while changing $\lambda'$
   in the range $0 \le \lambda' \le \lambda$.
   Here, the occupations are 
    $p_1=p_2=p_3=\frac{\lambda}{3 \lambda + \lambda'}$
   and $p_4= \frac{\lambda'}{3 \lambda + \lambda'}$,
and the amplitudes are,
       $a_1=-d$, $a_2=a_3=0$ and $a_4=d$.
The dephasing rates for weak coupling ($d \ll \lambda $) are
\begin{equation}
     \label{eqn:T2INT}
    T_2^{-1} = \left\{
    \begin{array}{rl}
    \frac{d^2}{4 \lambda},& \quad \mbox{for $\lambda'=\lambda$} \\
    \frac{ 2 d^2}{27 \lambda},& \quad \mbox{for $\lambda'=0$ .} \\
    \end{array}\right.
\end{equation}
In Fig. 2, the results
     by numerically solving Eq. (\ref{eqn:de}) are plotted.
This plot  shows that two limits ($T_2^{-1} = \frac{d^2}{4 \lambda}$
       and $\frac{2 d^2}{27 \lambda}$ )
       are smoothly connected in the intermediate parameter region
       for $d / \lambda =0.1$ (weak coupling case).
In the limit of no interaction
       ($\lambda=\lambda'$),
       the dephasing rate becomes twice of that 
       for the single charge trap.
For a strong interacting limit ($\lambda'=0$),
       the time evolution reduces to that of a single charge trap with
       asymmetric transition rate of 
       $\lambda_u^{single}=\frac{1}{3} \lambda_1$
       and $\lambda_d^{single}= \frac{2}{3} \lambda_1$ 
       with $\lambda_1 = \frac{2}{3} \lambda$
       and the dephasing rate is smaller than that of a single charge trap.
For $d / \lambda =2$, there is a rapid increases in the dephasing rate
        when $\lambda \sim \lambda'$.
There are four eigenvalues for characteristic equation of $R_{im}$.
Therefore there are four characteristic dephasing rates for this case.
This singularity appears because two of them become same,
      thus the transition from weak coupling to strong coupling occurs there.
Such a singularity also appears for the dephasing due to single charge trap.
\cite{Itakura_Tokura}                
All plots show that the 
       dephasing rate  increases with $\lambda'/\lambda$,
       which indicates that the effect of interaction, 
       or, the screening effect,
         suppresses the dephasing
       compared with that of non-interacting charge traps.
In this analysis, the average, $\eta$,
 changes with the ratio $\lambda'/\lambda$.
If we choose $a_i$ such that the average of $\eta$ is invariant,
  the results are the same.       
The reason is that for both cases,
$a_2 - a_1 = a_3 -a_1=d$ and $a_4 -a_2 =a_4 - a_3=d$,
independent of $\lambda'/\lambda$,
  and the difference between the former case and the 
  latter case only leads to the
    modulation of Rabi oscillation frequency.
    
We also examined the Gaussian behavior, which is the short-time regime
        for $t < min ( \frac{1}{d}, 3/(\max ( \lambda,\lambda' )))$
        \cite{Itakura_Tokura}.
For dephasing due to a single charge trap,
  the off-diagonal element of density matrix decay is represented as,
  $R (t) \simeq {\rm exp} ( - \frac{1}{2} ( \frac{t}{T_{2g}} )^2)$,
  where $T_{2g}^{-2}$ is given by $\frac{d^2}{4}$.
For two non-interacting charge traps,
  $T_{2g}^{-2}=\frac{d^2}{2}$,
  and for strong  interaction $(\lambda'=0)$,
  $T_{2g}^{-2} = \frac{ d^2 }{4}$.
This behavior shows that 
 the dephasing is suppressed as interaction increases,
 even for the Gaussian behavior.
It should be noted that the decay rate of Gaussian behavior 
  depends on the total charge of charge traps, 
  where we chose zero as the mean of amplitudes.
In present examinations, the effect of interaction 
  between charge traps is discussed,
  while          
  the numerical estimation of dephasing rate due to non-interacting BCF
         had been done in refs. \citen{Itakura_Tokura,Nakamura_CE,Fazio}.
Please note that         
         $T_{2g}^{-2}$ depends only on 
         distribution of $d^2$ and number of charge traps,
         although $T_{2}^{-1}$ depends on distributions of $d^2/\lambda$
         and number of charge traps; Eq. (\ref{eqn:T2INT}).
The former result coincides with that of ref. \citen{Fazio}
         when the initial state is in thermal equilibrium.

\begin{figure}
\center
\includegraphics[width=.55\linewidth]{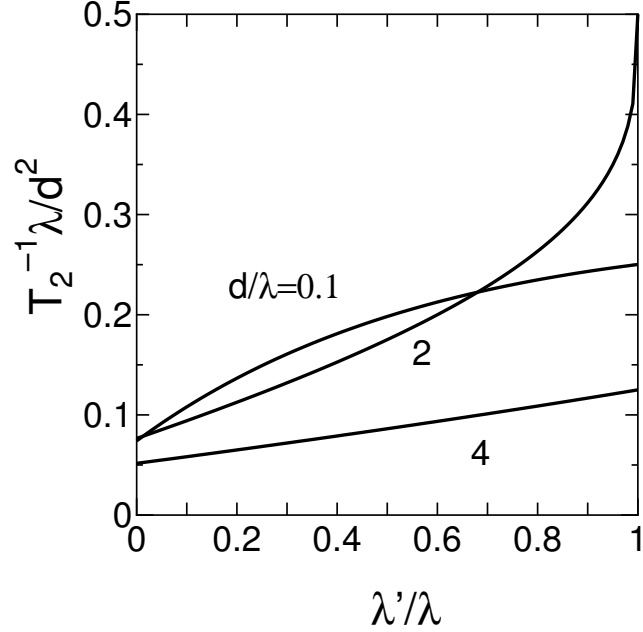}
\begin{minipage}[t]{8.5cm}
\caption{
The $\lambda'/\lambda$ dependence of dephasing rate 
   $T_{2}^{-1} \lambda/d^2$.
 The lines indicate the numerical results where the parameters are
 $d / \lambda =$ 0.1, 2 and 4.
}
\label{figure2}
\end{minipage}
\end{figure}

At lower temperatures than the energy differences of each state,
  we have asymmetric transition rates.
Therefore, we must consider the effect of temperature.
In order to satisfy the detailed balance condition,
 temperature and electron correlation leads to the following
        forms of 
        the transition rate
        \cite{Imry}:
$       \lambda_d = \lambda_d' = \lambda,
        \lambda_u=\lambda e^{-\frac{\Delta}{k_B T}}$
       and 
       $ \lambda_u'=\lambda e^{-\frac{\Delta+E_{charge}}{k_B T}}.$
The definitions of energy difference are: 
$\Delta=E_2-E_1=E_3-E_1$, $\Delta + E_{charge} = E_4 -E_2 = E_4 - E_3$.
The $E_{charge}$ is the capacitive energy between two charge traps.
In this case, the probabilities of population obey classical
      Boltzmann distribution, where
$      p_i = \frac{e^{-E_i/k_B T}}{\sum_{j=1}^4 e^{-E_j/k_B T}}.$
At high temperatures ($T \gg E_4/k_B $), 
      the occupation probabilities become $p_1=p_2=p_3=p_4=1/4$.
In Fig. 3, we show the dephasing rate obtained numerically
              with the amplitudes set to $a_1=-d,a_2=a_3=0,a_4=d$.
We chose the numerical parameter of $d/\lambda=0.1$
       (weak coupling).
\begin{figure}
\center
\includegraphics[width=.55\linewidth]{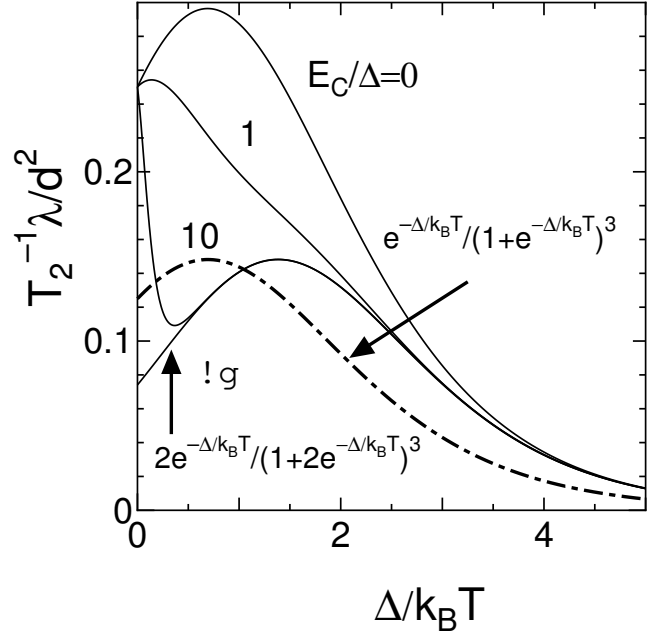}
\begin{minipage}[t]{8.5cm}
\caption{
The $\Delta/k_B T$ dependences of dephasing rate 
   $T_{2}^{-1} \lambda/d^2$
   with $d/\lambda=0.1$.
 The lines indicate the numerical results where the parameters are
 $E_{charge} / \Delta =$ 0, 0.1, 10 and $\infty$.
The dot-dashed line indicates the analytical curve for single charge trap.
}
\label{figure3}
\end{minipage}
\end{figure}
Using analytical expressions of dephasing rate
       for single charge trap, Eq. (\ref{eqn:T2weak}),
       the equation of the normalized dephasing rate 
       for weak coupling case ($d \ll \lambda$) is given by    
$ T_{2,single}^{-1} =  
 \frac{d^2}{\lambda}    
 \frac{ e^{-\Delta/ k_B T}}{(1+ e^{-\Delta/k_B T})^3}.$
The behavior of the traps for N=2 requires detailed examination.
For weak interaction ($E_{charge} \ll \Delta $),
    the dephasing rate due to the two charge traps is
     twice that of the dephasing rate due to the  single charge traps.
From Eq. (\ref{eqn:T2weak}), the dephasing rate becomes suppressed
   as the asymmetry of the transition rates increases.
At low temperatures ($k_B T \ll \Delta$),
      the dephasing rate is suppressed exponentially,
 because the asymmetry of the transition rates increases 
 with a decrease in temperature.
At high temperatures,
   ($k_B T \gg \Delta $),
   the dephasing rate is again suppressed.
The reason is that,
 the characteristic transition rate ($\lambda_u +\lambda_d$),
 increases as the temperature increases.
When coupling between the qubit and charge traps is weak ($d \ll {\rm max} 
(\lambda_u, \lambda_d$)),
 the magnitude of 
 the fluctuations in the trace of a state 
 on the Bloch sphere decreases with increasing,
 $\lambda_u+\lambda_d$,
 \cite{Itakura_Tokura}
 hence the dephasing rate decreases as well.

Next, we must consider the behavior of the traps
  when Coulomb interaction is strong ($E_{charge} \gg \Delta$). 
Except for very high temperatures
  ($k_B T \gg E_{charge} + \Delta$),
  the dephasing rate due to two dynamical charge traps is, 
$T_2^{-1} \lambda / d^2 = 
 2 e^{-\Delta/ k_B T} / ( 1 + 2 e^{-\Delta/ k_B T})^3$.
Then, comparing with Eq. (\ref{eqn:T2weak}),
two charge traps are equivalent to a single charge trap 
   with asymmetric transition rate,
$\lambda_d = \lambda$, 
$\lambda_u = 2 e^{-\Delta / k_B T} \lambda $.
At intermediate temperatures
    ($E_{charge} \gg k_B T \gg \Delta $),
   the dephasing rate  is
   smaller than that of a single charge trap.
The reason for this behavior
    is that two traps 
    behave as a single charge trap
    with a larger characteristic transition rate,
    $\lambda(1+2 e^{-\Delta / k_B T})$,
  compared with that of  
  the single charge trap $\lambda(1+e^{-\Delta / k_B T} )$.
At low temperatures, ($\Delta, E_{charge} \ll k_B T$),
the dephasing rate decreases exponentially as the temperature decreases
   irrespective of $E_{charge}$.   

Finally, we examine N identical charge traps which are located close to one
     another.
To simplify the discussion, we consider 
 a system of N charge traps symmetrically  coupled with a qubit.
There are: one empty state $(i=0)$, 
        N single occupied states $(i=1)$, (N-1)N/2 two occupied states
        $(i=2)$. $\cdots$, one fully occupied state $(i=N)$.
For strong and long-range 
     Coulomb interaction, the empty state and single occupied states 
     are relevant.
When there is 
     weak coupling, we have 
    $T_2^{-1} = \frac{d^2}{\lambda} 
    \frac{N e^{-\Delta/k_B T}}{(1+N e^{-\Delta/k_B T})^3}$,
    where $d$ and $\lambda$ are 
    coupling constants between the qubit and the background charges,
     and characteristic transition rate of charge traps,
     respectively,
     which are identical for all charge traps.
When there is non-interaction,
 N charge traps  behave  independently,
     and $T_2^{-1}  = 
    \frac{d^2}{\lambda} \frac{N e^{-\Delta/k_B T}}{(1+e^{-\Delta/k_B T})^3}$,
     hence the interaction between charge traps  suppresses 
     the dephasing rate.
At high temperatures, the analytical solution of 
$T_2^{-1}$ and $T_{2g}^{-2}$ are given by
$  \frac{T_2^{-1} ( {\rm strong \quad interaction} ) }{T_2^{-1} ( {\rm free} )}
  = \frac{8}{(N+1)^3}, $  
$  \frac{T_{2g}^{-2} ( {\rm strong \quad interaction} )}
  {T_{2g}^{-2} ( {\rm free} )}
  = \frac{1}{N},$ 
  where we chose zero as the mean of amplitudes.
Therefore, the dephasing rate becomes suppressed more effectively
 as the number of charge traps increases.
It should be noted that when charge traps are interacting strongly each other,
       the dephasing rate with dissipation in the large N limit
       is given by $\frac{E_J^2}{\delta E_C^2 + E_J^2} 
       \frac{d^2}{N \lambda}$.
Then the dephasing rate with dissipation becomes also suppressed 
       with increasing  N.
While the dephasing rate with dissipation becomes gradually dominant
       over the pure dephasing rate as increasing N,
       we do not argue this effect since both
       rate vanish with N.

In conclusion, we examined the dephasing rate of a two-level system,
        coupled with a classical environment made of N charge traps.
The environment changes its bistable extra bias,
        which results in pure dephasing.
When the charge traps fluctuate independently,
        the total dephasing rate is the simple summation of
        the dephasing rate of each charge trap.
If multiple charge traps are interacting with each other,
       the dephasing rate is slowed,
       when $T$ is not much smaller than $\Delta/k_B$.
At high temperatures, $(T \gg \Delta/k_B )$,
         more than one charge traps with 
         large Coulomb interaction results in a smaller dephasing rate
         than that of the single charge trap.
It should be noted that the other channels of dephasing exist,
         such a dephasing rate should be added to present dephasing rate.
And present estimation of dephasing rates corresponds to
        that of free induction decay, 
        \cite{Nakamura_CE}
        not that during gate operation in such a case
        the charge degeneracy state ($\delta E_C =0$)
        should be manipulated.
The numerical evaluation of dephasing rate
        for such a situation has been done in
        refs. \citen{Itakura_Tokura,Nakamura_CE}.

{\bf Acknowledgements} 
The authors thank Toshimasa Fujisawa, Yoshiro Hirayama, 
Gerrit E. W. Bauer and Fumitada Itakura 
for their advice and stimulating discussions.
This work was partly supported by CREST-JST.


\end{document}